\documentclass[journal]{IEEEtran}
\usepackage{graphicx}
\usepackage{epsf}
\usepackage{mathtools}
\usepackage{amssymb}
\usepackage{amsmath, amsfonts}
\usepackage{latexsym}
\usepackage{multirow}
\usepackage{multicol}
\usepackage[table]{xcolor}
\usepackage{color}

\usepackage{tabularx}
\usepackage{lipsum}

\usepackage{mathrsfs}

\usepackage{fixltx2e}

\usepackage[mathscr]{euscript}

\usepackage{commath}
\usepackage{MnSymbol}

\usepackage{subcaption}

\usepackage{mathtools} 
\DeclarePairedDelimiterX\setc[2]{[}{]}{\,#1 \;\delimsize\vert\; #2\,}
\DeclarePairedDelimiterX\parth[2]{(}{)}{\,#1 \;\delimsize\vert\; #2\,}

\DeclareMathOperator*{\argmax}{argmax}

\DeclarePairedDelimiter{\floor}{\lfloor}{\rfloor}

\PassOptionsToPackage{hyphens}{url}
\usepackage[unicode=true, bookmarks=true, bookmarksnumbered=true, bookmarksopen=true, bookmarksopenlevel=1, breaklinks=false, pdfborder={0 0 0}, pdfborderstyle={}, backref=false, colorlinks=false]{hyperref}

\usepackage[linesnumbered,ruled,vlined]{algorithm2e}

\usepackage{tikz}

\usepackage{theorem}
{
\theorembodyfont{\rmfamily}

}

\definecolor{orange}{RGB}{255,127,0}
\definecolor{blue}{RGB}{0,0,255}
\definecolor{red}{RGB}{255,0,0}
\definecolor{green}{RGB}{50,160,50}
\definecolor{grey}{RGB}{125,120,125}
\definecolor{purple}{RGB}{125,0,125}

\begin{document}
{
\title{{\fontsize{18}{2}\selectfont Environment-Adaptive Multiple Access for Distributed V2X Network:\\\vspace{-0.09 in}A Reinforcement Learning Framework}}

\author
{
Seungmo Kim, \textit{Member}, \textit{IEEE}, Byung-Jun Kim, and B. Brian Park, \textit{Senior Member}, \textit{IEEE}

\thanks{S. Kim is with the Department of Electrical and Computer Engineering, Georgia Southern University in Statesboro, GA. B. J. Kim is with the Department of Mathematical Sciences, Michigan Technological University in Houghton, MI. B. B. Park is with the Link Lab \& Department of Engineering Systems and Environment, University of Virginia, Charlottesville, VA. The corresponding author is S. Kim who can be reached at seungmokim@georgiasouthern.edu.

This work was supported in part by the Georgia Department of Transportation (GDOT) via grant RP 20-03.}

\vspace{-0.3 in}
}

\maketitle
\begin{abstract}
Cellular vehicle-to-everything (C-V2X) communications amass research interest in recent days because of its ability to schedule multiple access more efficiently as compared to its predecessor technology, i.e., dedicated short-range communications (DSRC). However, the foremost issue still remains: a vehicle needs to keep the V2X performance in a highly dynamic environment. This paper proposes a way to exploit the dynamicity. That is, we propose a resource allocation mechanism adaptive to the environment, which can be an efficient solution for air interface congestion that a V2X network often suffers from. Specifically, the proposed mechanism aims at granting a higher chance of transmission to a vehicle with a higher crash risk. As such, the channel access is prioritized to those with urgent needs. The adaptation is implemented based on reinforcement learning (RL). We model the RL framework as a contextual multi-armed bandit (MAB), which provides efficiency as well as accuracy. We highlight the most superb aspect of the proposed mechanism: it is designed to be operated at a vehicle autonomously without need for any assistance from a central entity. Henceforth, the proposed framework is expected to make a particular fit to distributed V2X network such as C-V2X mode 4.
\end{abstract}


\section{Introduction}\label{sec_intro}

\subsubsection{Background}
It is no secrete any more that vehicle-to-everything (V2X) communications hold massive potential for realizing intelligent transportation system (ITS). Nonetheless, at the same time, we encounter various technical challenges in deploying V2X communications in practice.

Especially in the United States (U.S.), the decision on a long debate on the 5.9 GHz band (i.e., 5.850-5.925 GHz) came out that the lower 45 MHz will be taken by Wi-Fi (including outdoor operations allowed \cite{lett}) while the ITS operations will be kept in the upper 30 MHz. Furthermore, the U.S. Federal Communications Commission (FCC) decided to oust dedicated short-range communications (DSRC) \cite{bennis19}, the long-time primary system of the band, while cellular V2X (C-V2X) will act as the technology with an exclusive right to operate ITS applications in the band \cite{nprm}.

As such, the ruling has now cleared debates on coexistence among the dissimilar systems \cite{globecom18} and has led to urgent need for thorough study on C-V2X. The technology started to adopt some smart methods in its multiple access across the physical (PHY) and the medium access control (MAC) layers. For instance, Long Term Evolution V2X (LTE-V2X) adopted the demodulation reference signal (DMRS) density is increased as an effort to enable a vehicle to efficiently perform the channel estimation and synchronization tracing even in high Doppler cases with a very high speed \cite{cv2xchina}. Not only that, the LTE-V2X used turbo codes, hybrid automatic repeat request (HARQ), and single carrier frequency division multiplexing access (SC-FDM) as a means to achieve higher reliability. With the synchronous scheme and frequency division multiplexing (FDM) in resource allocation scheme of LTE-V2X, the spectral efficiency and the system capacity can be improved.

Lately, the impetus of evolution has got even more rapid with the introduction of 5G \cite{jsac}. While being complementary to its predecessor LTE-V2X, the 5G's version of V2X--namely, New Radio V2X (NR-V2X)--further evolved the PHY layer structure of sidelink signals, channels, bandwidth parts, and resource pools in such a way to support a wider variety of transmission types (i.e., unicast and groupcast) with available feedback besides broadcast.

However, there still remain issues to solve. In particular, due to high mobility and dynamicity \cite{access19}, it makes a compelling case to lighten communications load in C-V2X for minimizing latency and maximizing reliability. While some methods of lightening networking load for DSRC (such as \cite{milcom19}) have been introduced, C-V2X is still leaving much to explore possibly due to higher complexity in its resource management and scheduling mechanisms as compared to DSRC.

Interestingly, a vehicular network features a unique characteristic that each vehicle experiences an ever-changing environment due to the nature of mobility. We propose to exploit the environment as the main driver to coordinate multiple access in a V2X network. This makes a compelling case of proposing a reinforcement learning (RL)-based approach where a vehicle autonomously enriches knowledge about the environment over time and updates its V2X networking parameters on the fly.

To this end, this paper is positioned to be the first proposal of a RL framework aiming at lightening the load of a C-V2X network. Specifically, we propose a RL mechanism that optimizes the transport block size (TBS) according to environment that a vehicle experiences. The proposed mechanism features its ability to be executed at each vehicle autonomously without any support from central entity. It yields that the proposed scheme can be particularly useful in the distributed mode (i.e., mode 4) of a C-V2X network, which has been regarded a challenging type of system to manage multiple access as compared to mode 3.

\subsubsection{Related Work}
In the literature, several \textit{learning-based} resource allocation methods for V2X network have been proposed. One main body of the prior work is RL. Compared to other methods (i.e, supervised and unsupervised learning \cite{LiY18}), RL has received increasing attention in solving difficult adaptation problems \cite{YeL19}\cite{iet20}, thanks to its ability to treat environment dynamics in a sequential manner \cite{SuB98}. However, feature representation and online learning ability are two major challenges to be solved for learning control of uncertain dynamic systems \cite{LiH20}. As an effort to keep a V2X network's performance stable in such a dynamic environment, a recent work \cite{fabric20} has proposed to apply a MAB-based approach, which turned out to be effective in achieving convergence of learning in a sufficiently short time to deal with the dynamicity. Meanwhile, advanced methods such as federated learning has recently been proposed as a solution to achieve self-adaptation of a wireless system \cite{NgP20}; however, its ``localized'' validity does not suit our goal of achieving a universal finality.

Moreover, as a method dealing with the time variance of the input in a RL framework, online learning enables adaptations with data being available in a ``streaming'' manner, as opposed to the offline learning that is trained by an entire training data set at once \cite{Sli19}\cite{ZhP20}. As such, the technique is known to be particularly efficient in areas where it is computationally infeasible to train over the entire dataset.

Distinguished from the above-viewed prior work, this paper targets to improve such current setting in such a way that the C-V2X can differentiate the priority of access according to the level of danger. Specifically, this paper finds quantification of environmental state of a vehicle particularly challenging due to its spatiotemporal dynamicity. In that regard, prior to this paper, the authors have been building a similar framework \cite{secon19}-\cite{access20}. This work is a significant extension of them in the sense that this work designs a RL framework with the input of driver behaviors, while the prior work focused on external factors. (Considering the level of recent onboard sensor technologies \cite{park20}, it is plausible to posit that the driver behaviors can be detected at an acceptable accuracy.) Another key improvement is that this work proposes to design C-V2X while the previous work discussed DSRC.

\begin{table}[t]
\centering
\caption{Frequently used symbols and acronyms}
\label{table_definitions}
\begin{tabular}{ll}
\hline
{\cellcolor{gray!20}{\textbf{Label}}} & {\cellcolor{gray!20}{\textbf{Definition}}} \\ \hline
$\left(\alpha,\beta\right)$ & Beta distribution parameters indicating a (success, failure)\\
BLER & Block error rate\\
HARQ & Hybrid automatic repeat request\\
MAB & Multi-armed bandit\\
NPRB & Number of physical resource blocks\\
NR-V2X & New Radio vehicle-to-everything\\
PSFCH & Physical sidelink feedback channel\\
$r$ & Reward\\
RL & Reinforcement learning\\
SL-SCH & Sidelink shared channel\\
TBS & Transport block size\\
$\mathbf{x}_{i}$ & Vector containing driver behavior types (w/ size $N \times 1$)\\
$x_{j}$ & $\mathbf{x}_{i}$'s $j$th element, denoting a behavior type $i$\\
$\mathbf{y}$ & Action vector (w/ size $M \times 1$)\\
$y$ & A value of the action, denoting a TBS value\\ \hline
\end{tabular}
\end{table}

\subsubsection{Contributions of This Paper}
Being uniquely positioned to extend the current literature as aforementioned, this paper highlights several technical contributions:
\begin{itemize}
\item It provides a framework of \textit{quantifying the crash risk} around a vehicle;
\item It presents a RL algorithm that optimizes the resource allocation for sidelink communications in NR-V2X mode 4, adaptive to the quantified crash risk;
\item The RL algorithm itself features \textit{autonomous} operation at a vehicle without need for any support from a centralized entity (e.g., server or network core)
\end{itemize}

\begin{figure*}
\centering
\includegraphics[width = 0.85\linewidth]{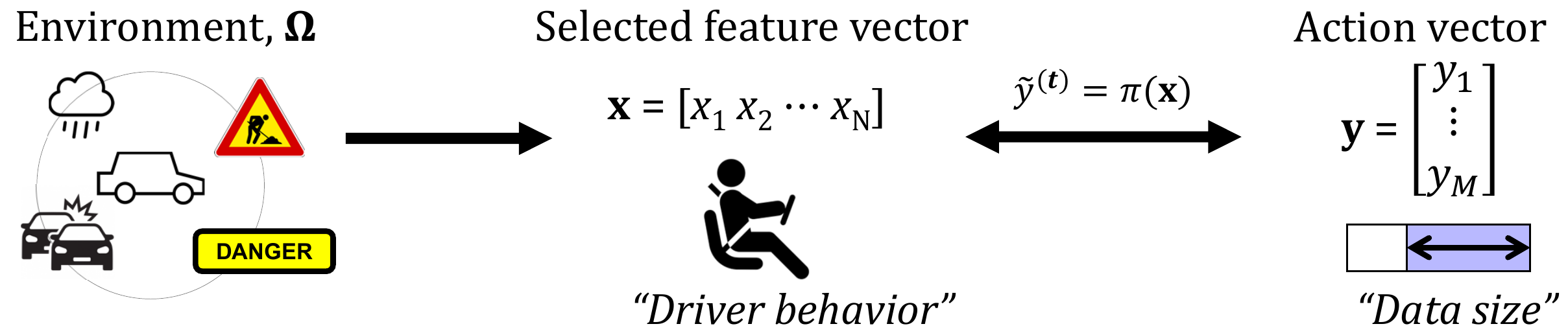}
\caption{Overview of the problem formulation and solving method ($\hat{\textbf{y}}^{(t)}$: A set of action values at time $t$; $\pi$: The policy of selecting an action given a state of the vehicle)}
\label{fig_overview}
\end{figure*}

\section{System Model: 3GPP NR-V2X Mode 4}\label{sec_model}
This paper postulates the connection type of a network to be completely \textit{distributed}. As such, the model naturally applies to C-V2X mode 4 where the nodes are connected directly in a distributed manner without going through the network core. In what follows, we spell out key technical details defining PHY and MAC layers of the 3GPP NR-V2X.

\subsubsection{Sidelink}\label{sec_model_sidelink}
The 3GPP introduced \textit{sidelink} in Release 12 as the third option after downlink and uplink mainly for the support of device-to-device communications. As the standardization organization introduced LTE-V2X in Release 14, the sidelink started to take a vital technical basis in supporting both basic safety and advanced use cases for ITS.

While being backward-compatible to the LTE-V2X, NR-V2X features some key technical enhancements. One example is the waveform type. Enhanced from LTE-V2X that uses single-carrier frequency-division multiple access (SC-FDMA), NR-V2X sidelink uses the cyclic-prefix orthogonal frequency division multiplexing (CP-OFDM) waveform with supporting multiple options for subcarrier spacings (i.e., 15, 30, 60 and 120 kHz) and modulation schemes (i.e., quadrature phase shift keying (QPSK), 16-quadrature amplitude modulation (QAM), 64-QAM, and 256-QAM).

\subsubsection{PSFCH}\label{sec_model_psfch}
It is significant to note that starting from Release 16, the NR-V2X adopted \textit{feedback} functions via the physical sidelink feedback channel (PSFCH) for unicast and groupcast \cite{tr37985}.

The PSFCH carries HARQ feedback over sidelink from a recipient (Rx) vehicle of a message over PHY sidelink shared channel (PSSCH). Sidelink HARQ feedback may be in two particular forms: (i) conventional acknowledgement (ACK)/negative acknowledgement (NACK); or (ii) NACK-only, i.e., nothing transmitted in case of successful decoding. (See Section 6.2.4. of \cite{tr37985})

We reiterate the significance of existence of such a feedback functionality since this paper proposes a RL framework, which essentially necessitates feedback (i.e., reward) as a result of an action.

\subsubsection{SPS}\label{sec_model_sps}
C-V2X mode 4 communication relies on a distributed resource allocation scheme, namely sensing-based \textit{semipersistent scheduling (SPS)} \cite{ts36213} which schedules radio resources in a standalone fashion at a vehicle. Owing to the characteristic of traffic that usually is periodic, it has been found effective to sense congestion on a resource and estimate a future congestion on the resource \cite{pimrc08}. Specifically, this estimation forms the basis on how the resource is booked.

In that way, the SPS minimizes the chance of ``double booking'' between transmitters that are using overlapping resources. To elaborate the technical details, a vehicle reserves certain resource blocks (RBs) for a random number of consecutive packets. This number depends on the number of packets transmitted per second, or inversely the packet transmission interval. As such, via a sidelink control information (SCI), each vehicle sends information its packet transmission interval and its reselection counter. Neighboring vehicles use this information to estimate which RBs are free when making their own reservation to reduce packet collisions. It leads to that vehicles autonomously select their resources without the assistance from the cellular infrastructure.

\subsubsection{Receiver}\label{sec_model_rx}
After receiving a signal, the first step that the Rx performs is synchronization. Then, the synchronized signals are passed to the CP-OFDM demodulation. It is followed by extraction of the DMRSs for channel estimation. (Notice that we do not assume perfect channel estimation for realistic modeling.) Now, the process turns into extraction of the data on the desired transport blocks (TBs). The resource allocation information is obtained from the corresponding SCI messages, which is always SCI format 1 in V2X \cite{tr37985}. Then, equalization follows where we use a minimum mean square error (MMSE) equalizer. We do not formulate the channel and MMSE since the typical notations (i.e., $X$ for a transmitted signal, $H$ for a channel, $N$ for the complex white Gaussian noise with zero mean, and $Y$ for the received signal) conflict with other notations used in this paper (i.e., $x$ and $y$ for the input and output of the proposed RL loop).

\subsubsection{Performance Evaluation Metrics}\label{sec_model_metrics}
It is significant to notice that this paper relies on the \textit{block error rate (BLER)} and \textit{the normalized throughput} as metrics measuring the performance of a sidelink in NR-V2X.

First, the full definition of BLER can be found from one of the latest 3GPP technical specifications as the \textit{ratio of the number of erroneous blocks received to the total number of blocks sent}. An erroneous block is defined as a TB, the cyclic redundancy check (CRC) of which is wrong. (See Section F.6.1.1 of TS 34.121 \cite{ts34121}.)

Meanwhile, the normalized throughput is defined as
\begin{align}\label{throughput}
R = \frac{\mathsf{N}_{\text{bits, tx}}}{\mathsf{N}_{\text{bits, sf}} \times \floor{\mathsf{N}_{\text{sfs}}\hspace{0.02 in} / \hspace{0.02 in} \mathsf{N}_{\text{sfs, harq}}}}
\end{align}
where $\mathsf{N}_{\text{bits, tx}}$ gives the number of transmitted bits; $\mathsf{N}_{\text{bits, sf}}$ is the maximum number of bits that can be contained in a subframe; $\mathsf{N}_{\text{sfs}}$ gives the number of subframes that have been observed in a simulation; $\mathsf{N}_{\text{sfs, harq}}$ indicates the number of subframes between consecutive HARQ processes.

\section{Proposed Learning Mechanism for Optimal Sidelink Resource Allocation in C-V2X Mode 4}\label{sec_proposed}
We remind that the ultimate goal of our proposition is to design a resource allocation mechanism for NR-V2X mode 4, in which each vehicle optimizes its operation according to its environmental state. We also remind that this paper is proposing to define the state of a vehicle as the level of danger measured at the vehicle, as an effort to design a mechanism optimizing the operation of a vehicle adaptive to the danger that the vehicle marks. This section presents details on how we quantify the danger of a vehicle, which will form the basis for a learning mechanism that will be performed thereafter.

\begin{table*}[t]
\centering
\caption{An example of driver-related crash causing factors \cite{fars} for constitution of relationship between $\mathbf{x}$ and $\mathbf{y}$}
\label{table_X}
\begin{tabular}{|l|l|l|}
\hline
{\cellcolor{gray!20}{$\mathbf{x}$ = \textbf{Input value}}} & {\cellcolor{gray!20}{\textbf{Driver distraction type}}} & {\cellcolor{gray!20}{$\mathbf{y}$ = \textbf{TBS index}}} \\ \hline
$x_{1}$ & Driving too fast for conditions or in excess of posted limit & 1\\ \hline
$x_{2}$ & Under the influence of alcohol, drugs, or medication & \multirow{5}{*}{2}\\
$x_{3}$ & Failure to keep in proper lane & \\
$x_{4}$ & Failure to yield right of way & \\
$x_{5}$ & Distracted (e.g., phone, talking, eating, etc) & \\
$x_{6}$ & Overcorrecting / Oversteering & \\ \hline
$x_{7}$ & Failure to obey traffic signs, signals, or officers & \multirow{3}{*}{3}\\
$x_{8}$ & Erratic, reckless, careless, or negligent operation of vehicle & \\
$x_{9}$ & Swerving due to wind, slippery surface, object, etc & \\ \hline
$x_{10}$ & Vision obscured due to rain, snow, glare, lights, etc & \multirow{4}{*}{4}\\
$x_{11}$ & Driving on wrong way / side of road & \\
$x_{12}$ & Drowsy, asleep, fatigued, ill, or blackout & \\
$x_{13}$ & Improper turn & \\ \hline
\end{tabular}
\end{table*}

\begin{figure}
\centering
\includegraphics[width = 0.9\linewidth]{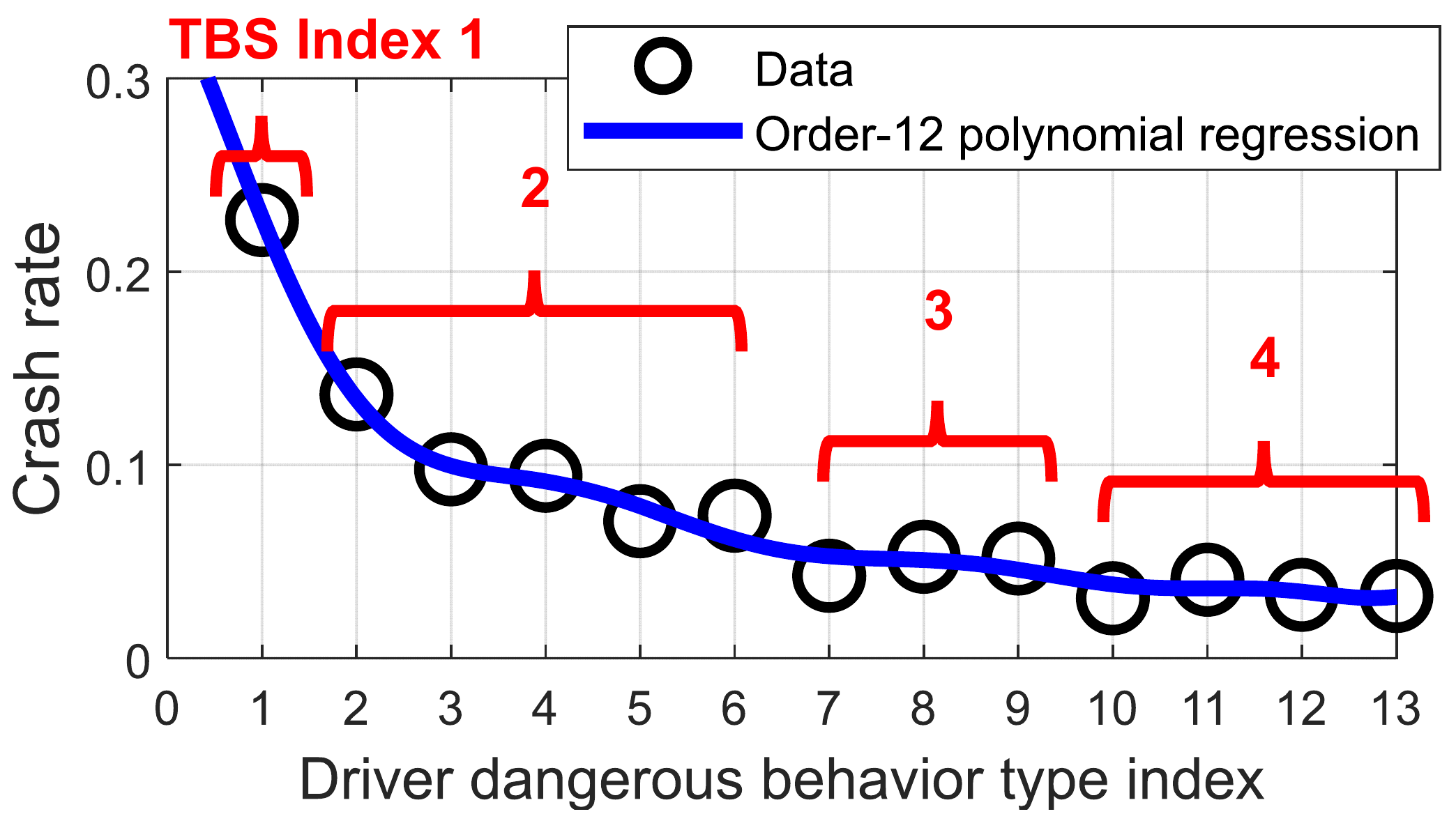}
\caption{Regression of driver's behavior type to TBS by using a 12-order polynomial $\mathbb{P}\left[\text{Crash}\right]_{x \in \mathbf{x}} = b_{1}x^{12} + b_{2}x^{11} + \cdots + b_{13}$ as an example of mapping $\mathbf{x}$ and $\mathbf{y}$ for the proposed RL mechanism}
\label{fig_order12poly}
\end{figure}

\subsection{Input Dimension Reduction}
Let the environment around a vehicle at time $t$ be denoted by $\mathbf{\Omega} \in \mathbb{R}^2$, which is composed of features defining the risk of a vehicle such as weather, vehicle speed, etc. As an important means to circumvent the curse of dimensionality, we map the large-volume space $\mathbf{\Omega}$ to a smaller space of a selected representative feature, i.e., an $N$-by-1 vector $\mathbf{x} := \left[x_{1} \hspace{0.07 in} x_{2} \hspace{0.07 in} \cdots \hspace{0.07 in} x_{N}\right]$ where each $x_{i}$ gives a value for the feature. Notice that the significance lies in \textit{what feature to extract} as a representative of the crash risk. In what follows, we elaborate the technical details on quantification of the environment, focusing on seeking answers to two key questions: \textbf{Q1:} From what dataset do we use to map $\mathbf{\Omega} \rightarrow \mathbf{x}$?; and \textbf{Q2}: Based on what rationale can we identify the feature $\mathbf{x}$?

Regarding \textbf{Q1}, we propose to draw from a nationwide dataset provided by the U.S. National Highway Traffic Safety Association (NHTSA) regarding fatal injuries suffered in motor vehicle traffic crashes, which is also known as the Fatality Analysis Reporting System (FARS) \cite{fars}. Let the entire FARS dataset be regarded $\mathbf{\Omega}$. Now, from $\mathbf{\Omega}$, we extract the most dominant feature $\mathbf{x}$, which serves as an estimate input with a reduced dimension.

Proceeding to addressing \textbf{Q2}, we identify the \textit{types of driver's dangerous behavior} as the key factor in defining the crash risk of a vehicle. More specifically, referring to the FARS dataset, we further identify key crash-causing driver behavior types in order to calculate $\mathbb{P}[\text{crash} \hspace{0.015 in} | \hspace{0.015 in} \mathbf{x}]$. As shown in Fig. \ref{fig_order12poly}, this probability provides criteria on which a C-V2X resource allocation mechanism is predicated on. To elaborate, a smaller TBS index is assigned for a highly crash-causing driver behavior type, which will yield a higher probability of successful message delivery and thus a higher chance of propagating the message to more vehicles in the network. This way, \textit{the air interface can be filled with more urgent messages} with a higher chance. It is also important to notice that the distribution shown in Fig. \ref{fig_order12poly} will be used as an initial factory setting for a vehicle, which will be updated in such a way that the distribution is customized over numerous drives according to the driver's behavioral characteristics while driving.

\subsection{Problem Formulation}\label{sec_proposed_formulation}
Now, we formulate a \textit{contextual MAB} between the context matrix $\mathbf{x}$ and a vehicle’s action $\mathbf{y} \hspace{0.015 in} | \hspace{0.015 in} \mathbf{x}$. That is, here we write a problem of finding an optimal policy, i.e., $\hat{\mathbf{y}}^{(t)} = \pi\left(\mathbf{x}\right)$ where $\hat{\mathbf{y}}^{(t)}$ denotes an action selected by the policy $\pi$ at time $t$.

Provided the relationship shown in Fig. \ref{fig_overview}, suppose a function $f$ mapping the original environmental space $\mathbf{\Omega}$ to the action space $\mathbf{y}$. Now, we note that the policy $\pi$ is an estimation of function $f$, due to the dimension reduction $\mathbf{\Omega} \rightarrow \mathbf{x}$. The key challenge here is that the selected feature $\mathbf{x}$ keeps updated in time $t$. As a means to deal with the challenge, we narrow our perspective down to \textit{establishing a RL mechanism autonomously updating the policy $\pi(\mathbf{x})$ based on time-varying $\mathbf{x}$.}

Henceforth, we translate the proposed environment-adaptive C-V2X resource allocation problem to a problem that finds an optimal policy selecting an optimal action given a context $\mathbf{x}$ at a given time $t$. We propose to formulate this problem as a variant of the \textit{0-1 knapsack problem (KP)} \cite{knapsack89} that aims to maximize the reward while keeping the cost under a certain level. Let the context at time $t$ be $\mathbf{x}^{(t)} = [x_{1}^{(t)} \hspace{0.015 in} \cdots \hspace{0.015 in} x_{N}^{(t)}] \in \mathbb{R}^{1 \times N}$ where $x_{i}^{(t)}$ gives the $i$th value of the feature $\mathbf{x}$. As has been illustrated in Fig. \ref{fig_overview}, we denote by $\mathbf{y} \in \mathbf{R}^{1 \times M}$, the vector of possible action values. We aim at keeping the problem as a finite-horizon decision problem, which means the optimal $\pi$ can be found within a finite number of time epochs. As such, modifying the KP, we formulate the process of predicting the optimal $\pi^{\ast}$, which is formally written as
\begin{align}\label{eq_KP}
\left(\mathbf{y}^{(t)}\right)^{\ast} &:= \pi^{\ast}\left(\mathbf{x}^{(t)}\right)\nonumber\\
&= \argmax_{y^{(t)} \in \mathbf{y}^{(t)}} \hspace{0.05 in} \sum_{k=1}^{K} r\left( y^{(t)} \hspace{0.015 in} | \hspace{0.015 in} x^{(t)} \right)\nonumber\\
&\hspace{-0.4 in} \text{s. t.} \displaystyle \sum_{y^{(t)} \in \mathbf{y}^{(t)}} c\left( y^{(t)} \hspace{0.015 in} | \hspace{0.015 in} x^{(t)} \right) \le C
\end{align}
where $K$ indicates the number of arms, i.e., number of TBS options. Also, $c(\cdot)$ denotes the cost and $C$ gives the maximum acceptable cost for operating action $y^{(t)}$ in context $x^{(t)}$.

As an important reminder, TBS represents the action space $\mathbf{y}$ in this paper, which is rationalized as follows. It is obvious that there are numerous factors determining the performance of a C-V2X system including TBS, modulation and coding scheme (MCS), DMRS density, waveform, OFDM numerology, and etc. (For instance, it is critical for OFDM to operate with an adequate set of parameters such as subcarrier spacing, number of slots per subframe, and slot length \cite{psun}.) We choose TBS since it makes the most plausible case that we control the \textit{payload size} according to the context related to the crash risk. In other words, a vehicle at a higher crash risk due to a dangerously behaving driver transmits a message with a smaller size so it can be delivered at a higher chance of success. See Section \ref{sec_results_TBS} for further details on our selection of TBS as $\mathbf{y}$ for the proposed learning mechanism.

\begin{algorithm}[hbtp]
\SetAlgoLined
\%--- \textit{Initial factory setting} ---\%\\ \vspace{-0.02 in}
\hspace{0.05 in} $\mathbf{x}^{(0)} \longleftarrow \mathbf{x}_{\text{ini}, N \times 1}$;\\ \vspace{-0.02 in}
\hspace{0.05 in} $\mathbf{y}^{(0)} \longleftarrow \mathbf{0}_{M \times 1}$;\\ \vspace{-0.02 in}
\hspace{0.05 in} $r^{(0)} \longleftarrow 0$;\\ \vspace{-0.02 in}
\vspace{0.07 in}

\For{t = 1, $\cdots$, $\infty$}{\vspace{0.03 in}

\%--- \textit{Input vector update} ---\%\\ \vspace{-0.02 in}
\If{Dangerous driver behavior detected}{
$\mathbf{x}^{(t)} \longleftarrow \mathbf{x}_{N \times 1}^{(t)}$;
}\vspace{0.07 in}

\%--- \textit{V2X for unicast or groupcast} ---\%\\ \vspace{-0.02 in}
\If{Received a msg to send from upper layer}{
\%--- \textit{Thompson sampling} ---\%\\ \vspace{-0.02 in}
Sample $\hat{\theta}_{k}^{(t)} \sim \text{Beta}\left(\alpha_{k}^{(t)},\beta_{k}^{(t)}\right)$ for $k = 1, \cdots, M$;\\ \vspace{-0.02 in}
Select arm $\hat{k}^{(t)} \longleftarrow \max_{k} \mathbf{\hat{\theta}}_{k}^{(t)}$;\\
Take action $\hat{y}^{(t)} \longleftarrow y|_{\hat{k}^{(t)}}$;\\
\vspace{0.1 in}
\% Observe reward\\ \vspace{-0.02 in}
\eIf{Correct TBS selection}{ \vspace{-0.04 in}
$r^{(t)} \longleftarrow 1$; \vspace{-0.04 in}
}{ \vspace{-0.04 in}
$r^{(t)} \longleftarrow 0$;
}
\vspace{0.07 in}
\% Update Beta distribution\\ \vspace{-0.02 in}
$\left(\alpha_{k}^{(t)},\hspace{0.01 in} \beta_{k}^{(t)}\right) \longleftarrow \Big(\alpha_{k}^{(t-1)} + r^{(t)},$\\
\vspace{-0.035 in} $\hspace{1.5 in}\beta_{k}^{(t-1)} + \left(1 - r^{(t)}\right) \Big)$;
}\vspace{0.03 in}

}
\caption{Proposed RL-based data size optimization algorithm at a vehicle for 5G NR-V2X mode 4 sidelink unicast and groupcast}
\label{algorithm_rl}
\end{algorithm}

\subsection{Problem Solving Algorithm}
Algorithm \ref{algorithm_rl} presents a pseuodocode for the proposed mechanism. We remind that the algorithm aims to learn an optimal TBS for a sidelink transmission (i.e., unicast or groupcast) in a NR-V2X mode 4 network.

Lines 1-4 indicate the initial setting of key variables. While initialization of $\mathbf{y}$ and $r$ are straightforward, that of $\mathbf{x}$ takes a bit further discussion. Let $\mathbf{x}_{\text{ini}}$ denote the initial distribution of $\mathbf{x}_{i}$ and be given to every vehicle as a factory setting. We recall that such a factory setting does not come out of the blue: an example of the setting can be founded on a nationwide consensus by a U.S. federal agency \cite{fars}, which has been discussed in Fig. \ref{fig_order12poly}. By $w_{j}$, we denote the \textit{weight} of the $j$th level of driver's dangerous behavior, which forms the Y-axis of Fig. \ref{fig_order12poly}. As such, the $\mathbf{x}_{\text{ini}}$ provides an initial mapping between $x_{j}$ and its weight $w_{j}$.

Through Lines 6-9, we recall that a vehicle is supposed to update this distribution reflecting its driver's driving behavior over time, which yields that the weights $w_{j}$ will be distributed differently according to (i) a time instant $t$ and (ii) vehicle index $i$. Specifically, the input vector $x_{j}$ is updated when the driver behaves differently from the initial setting $\mathbf{x}_{\text{ini}}$.

Lines 10-23 execute an event where the vehicle receives a message to send from the upper layer. We remind that this paper postulates a unicast or groupcast since they are the types of transmission providing \textit{feedback}, as per the latest 3GPP NR-V2X standard. (See Section 6.2.4 of \cite{tr37985}.

To break down, through Lines 12-14, the algorithm runs a TS wherein the vehicle (i) samples following the current Beta distribution and (ii) selects a TBS value according to the sampling. The algorithm proceeds to Lines 15-20 in which the vehicle observes the reward of the action. As written in Lines 21-23, the vehicle updates the Beta distribution based on the success and failure of the latest action. It is important to note that the reward is defined by \textit{whether the agent has selected a correct arm}, i.e., a correct TBS.

\begin{figure*}
\centering
\begin{subfigure}{0.49\linewidth}
\centering
\includegraphics[width = \linewidth]{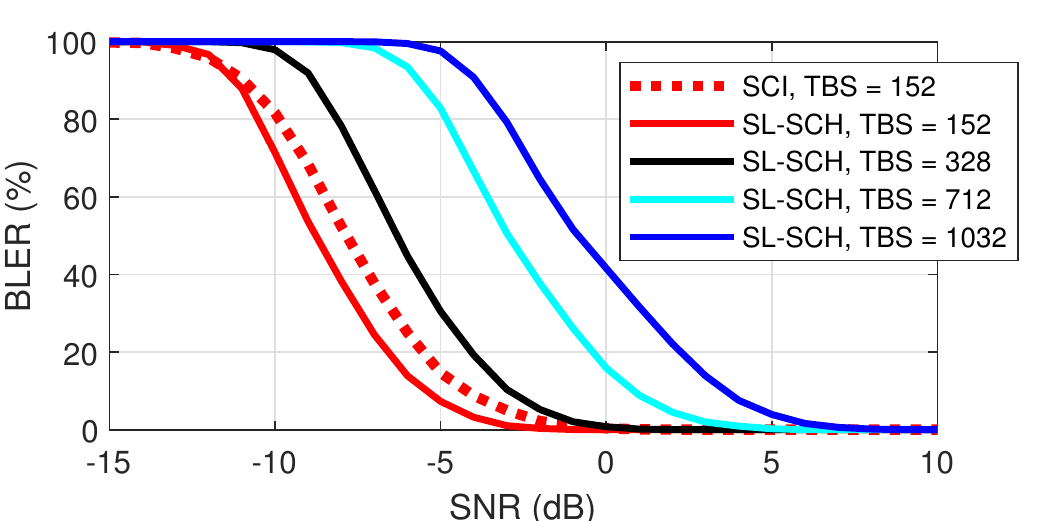}
\caption{BLER with NPRB = 6}
\label{fig_bler_nprb6}
\end{subfigure}
\begin{subfigure}{0.49\linewidth}
\centering
\includegraphics[width = \linewidth]{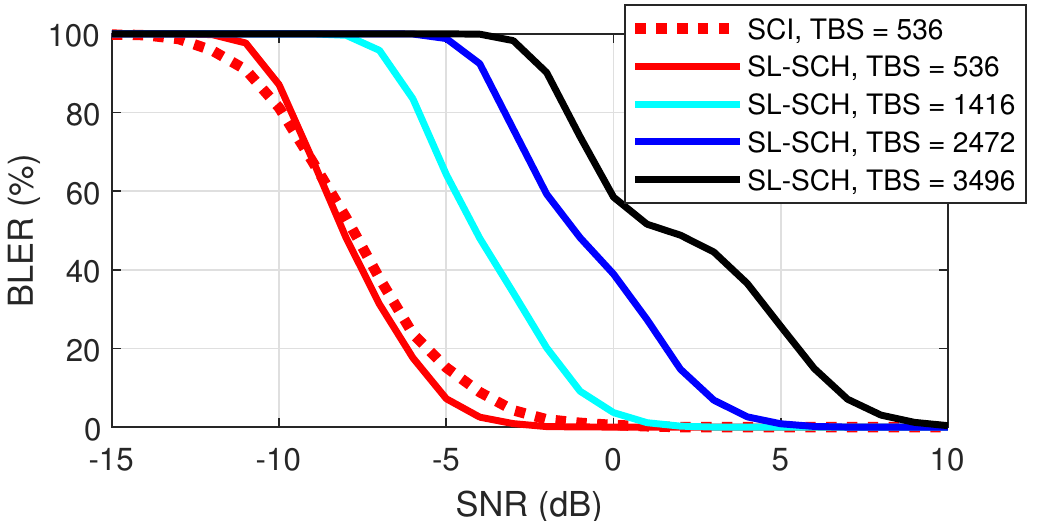}
\caption{BLER with NPRB = 20}
\label{fig_bler_nprb20}
\end{subfigure}
\begin{subfigure}{0.49\linewidth}
\centering
\includegraphics[width = \linewidth]{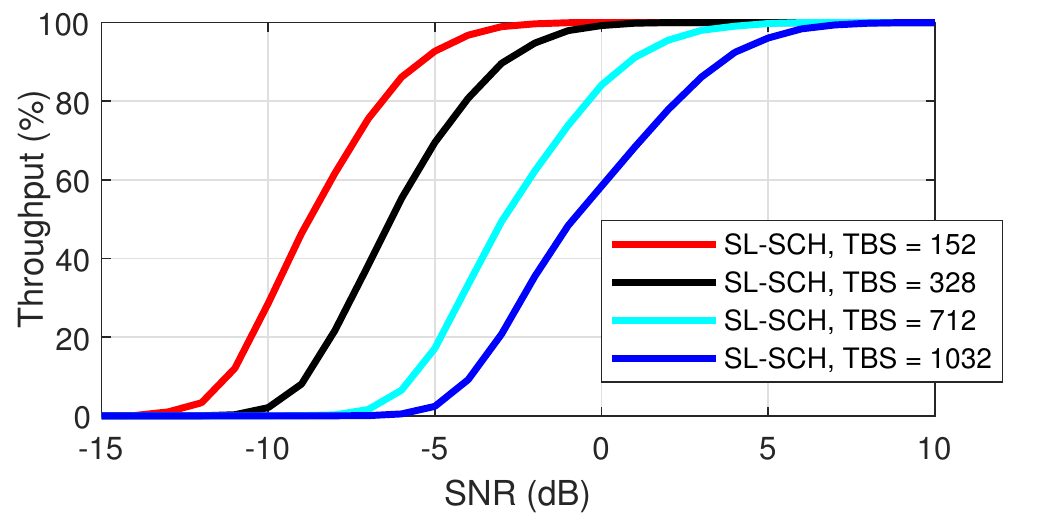}
\caption{Normalized throughput with NPRB = 6}
\label{fig_tput_nprb6}
\end{subfigure}
\begin{subfigure}{0.49\linewidth}
\centering
\includegraphics[width = \linewidth]{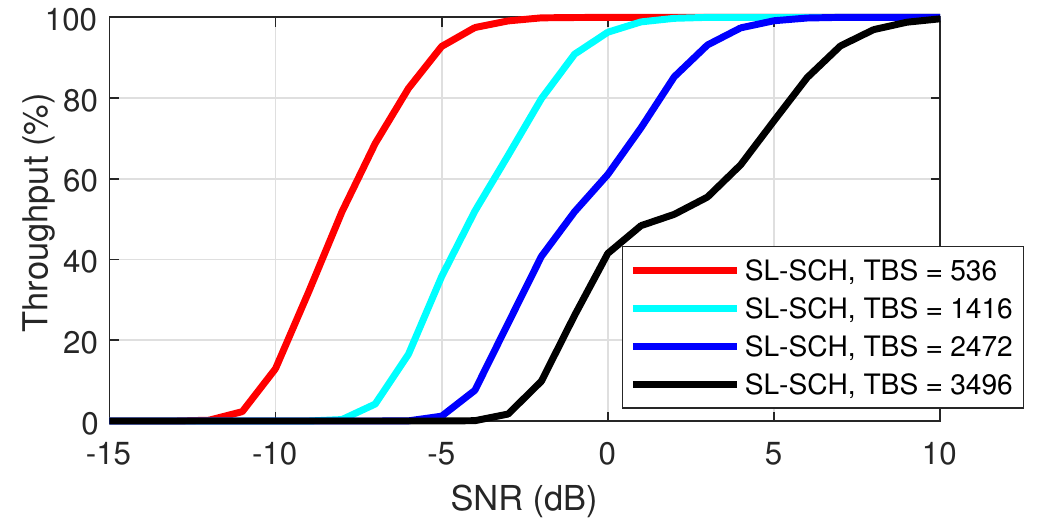}
\caption{Normalized throughput with NPRB = 20}
\label{fig_tput_nprb20}
\end{subfigure}
\caption{Performance of SL-SCH in NR-V2X mode 4 in terms of BLER and normalized throughput}
\end{figure*}

\section{Results and Discussions}\label{sec_results}
The baseline configuration is taken from the ``Reference measurement channel for transmitter characteristics'' as defined by Table A.8.3-1 \cite{ts36101}. Table \ref{table_parameters} summarizes the parameters. Notice that to simulate realistic V2X transmissions, multiple hybrid automatic retransmission request (HARQ) processes and retransmissions have been introduced in this simulation.

\subsection{Sidelink Performance according to TBS}\label{sec_results_TBS}
We start with corroborating that the TBS is a plausible factor to distinguish the performance of a NR-V2X network. Figs. \ref{fig_bler_nprb6} through \ref{fig_tput_nprb20} show the BLER and the normalized throughput versus SNR. The figures also demonstrate the performance being distinguished according to NPRB. Notice that we postulate four different options for the TBS. (See Section \ref{sec_results_slsch}) However, we stress that the framework is extendible: any other TBS value defined in Table 7.1.7.2.1-1 of TS 36.213 \cite{ts36213} could be eligible in the output space $\mathbf{y}$.

\begin{figure}
\centering
\begin{subfigure}{0.98\linewidth}
\centering
\includegraphics[width = \linewidth]{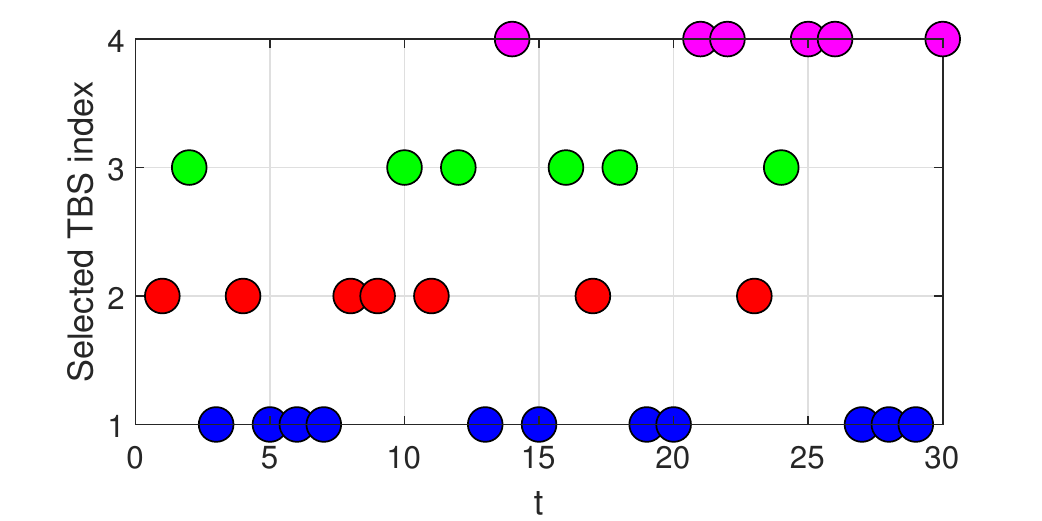}
\caption{Selection of TBS index: A/B testing}
\label{fig_TS_convergence_AB}
\end{subfigure}
\begin{subfigure}{0.98\linewidth}
\centering
\includegraphics[width = \linewidth]{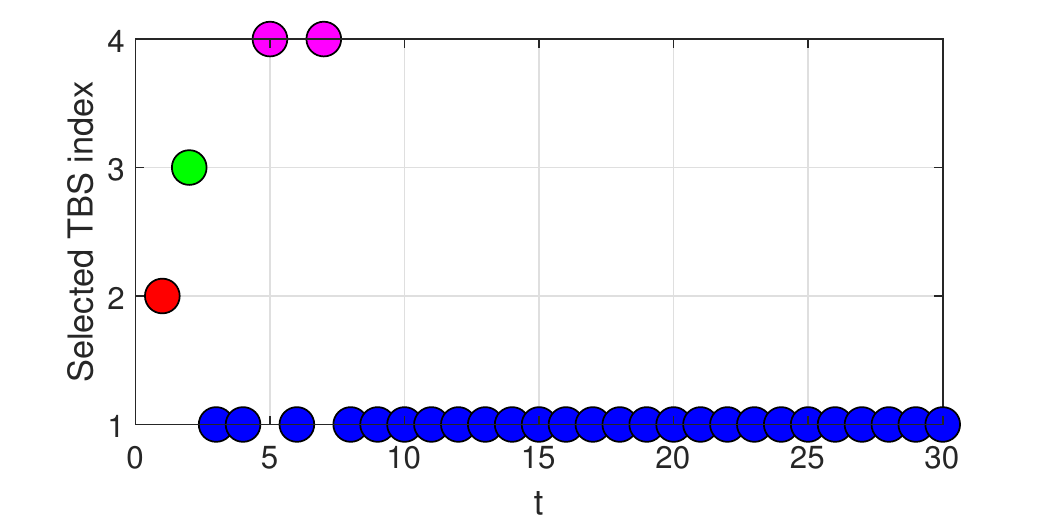}
\caption{Selection of TBS index: TS (Proposed)}
\label{fig_TS_convergence_TS}
\end{subfigure}
\begin{subfigure}{0.98\linewidth}
\centering
\includegraphics[width = \linewidth]{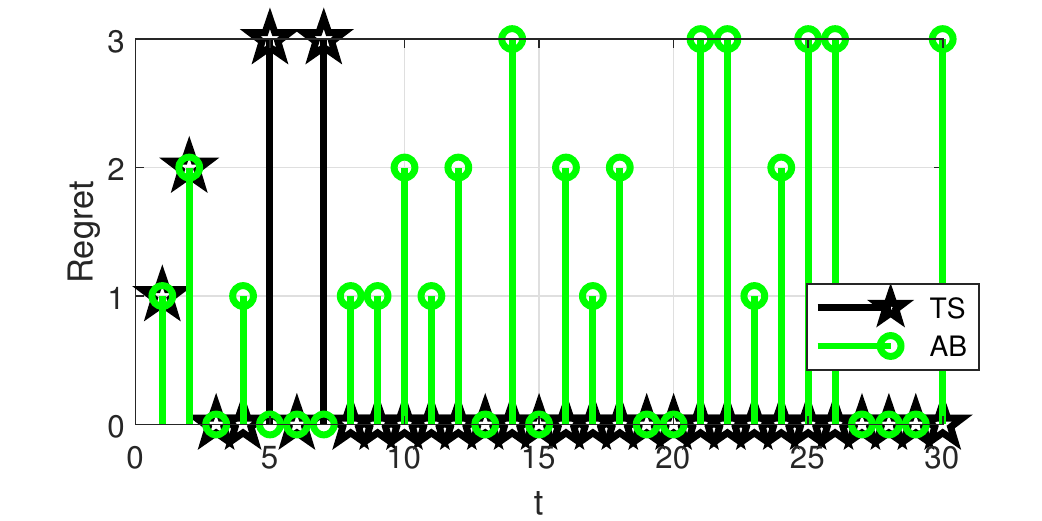}
\caption{Regret}
\label{fig_TS_regret}
\end{subfigure}
\caption{Example run of the proposed RL mechanism with assumption of $y^{\ast} = \pi\left(x_{1}\right) = 1$}
\label{fig_TS}
\end{figure}

\begin{table}[t]
\centering
\caption{Parameters \cite{ts36213}\cite{ts36101}}
\label{table_parameters}
\begin{tabular}{|c|c|}
\hline
{\cellcolor{gray!20}{\textbf{Parameter}}} & {\cellcolor{gray!20}{\textbf{Value}}} \\ \hline
System & 3GPP Release 16\\ \hline
Bandwidth & 10 MHz\\ \hline
Duplex mode & FDD\\ \hline
CP mode & Normal\\ \hline
Modulation & QPSK\\ \hline
\# Rx antennas & 2\\ \hline
Delay profile & Extended Vehicular A model (EVA) \cite{ts36104}\\ \hline
Doppler frequency & 500 Hz\\ \hline
Fading & Rayleigh\\ \hline
Equalization & MMSE\\ \hline
\end{tabular}
\end{table}

\subsection{Convergence and Accuracy of the Proposed RL Mechanism}
Fig. \ref{fig_TS} displays the average length of time taken for selection of the optimal TBS index for NR-V2X, as a means to evaluate the time complexity of the proposed RL scheme. Based on that we model the MAB problem as a Bernoulli-bandit, we evaluate the convergence performance based on TS. Over other algorithms to solve a MAB problem, TS has been evidenced to outperform other alternatives such as $\epsilon$-greedy and upper confidence bound (UCB) \cite{ms11}.

As an example, we set TBS index 1 as the \textit{successful} selection among the 4 different values for TBS as has been presented in Table \ref{table_X}. Comparison between Fig. \ref{fig_TS_convergence_AB} and \ref{fig_TS_convergence_TS} substantiates that the convergence of the proposed mechanism based on TS. We inform that this result is from 30 rounds of simulation where a vehicle learns on 4 arms representing the 4 TBS indices. One can observe from Fig. \ref{fig_TS_convergence_TS} that the proposed algorithm consumes first 7 runs on ``exploring'' the four arms as a means of training. The convergences of A/B testing and the proposed mechanism shown in Figs. \ref{fig_TS_convergence_AB} and \ref{fig_TS_convergence_TS} lead to the difference in terms of \textit{regret} as shown in Fig. \ref{fig_TS_regret}. Notice that the regret measured at time $t$ with arm $k$ selected is denoted by $\rho$, which is formally written as $\rho^{(t)} = \left|\left(y_{k}^{(t)}\right)^{\ast} - \hat{y}_{k}^{(t)}\right|$.

\begin{figure}
\centering
\begin{subfigure}{0.98\linewidth}
\centering
\includegraphics[width = \linewidth]{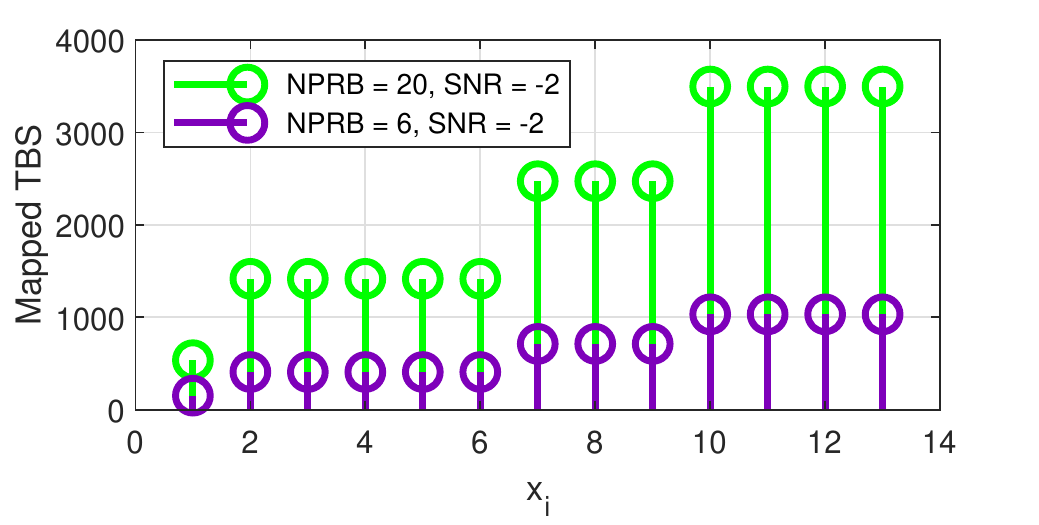}
\caption{TBS mapping versus $x_{i}$}
\label{fig_xi_mappedTBS}
\end{subfigure}
\begin{subfigure}{0.98\linewidth}
\centering
\includegraphics[width = \linewidth]{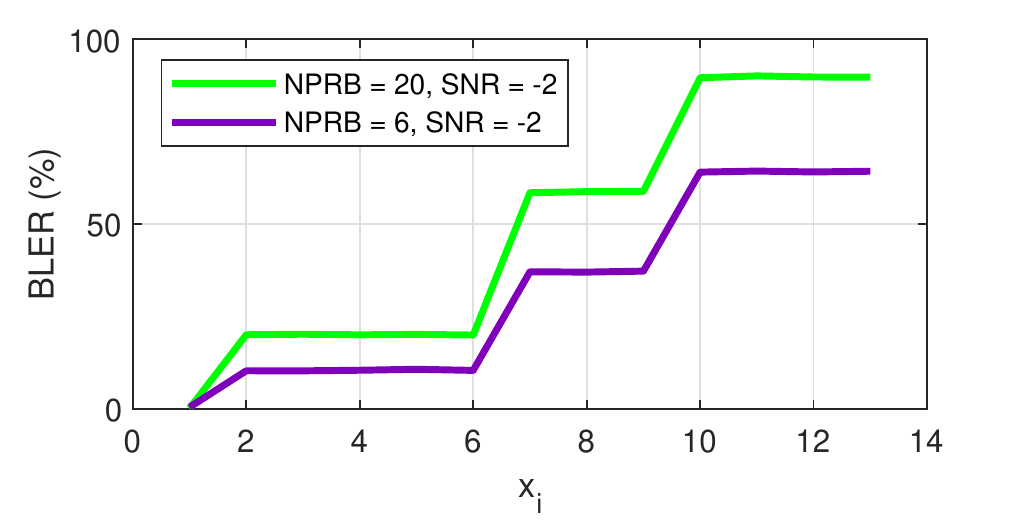}
\caption{Resulting BLER versus $x_{i}$}
\label{fig_xi_bler}
\end{subfigure}
\begin{subfigure}{0.98\linewidth}
\centering
\includegraphics[width = \linewidth]{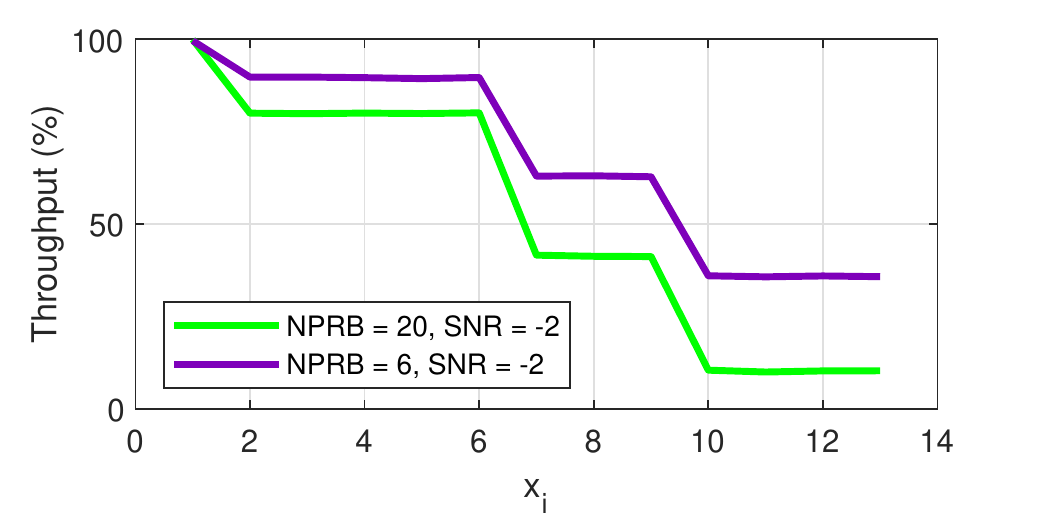}
\caption{Resulting throughput versus $x_{i}$}
\label{fig_xi_tput}
\end{subfigure}
\caption{(NPRB = \{6, 10\}, SNR = -2 dB)}
\label{fig_xi}
\end{figure}

\subsection{NR-V2X Performance with the Proposed Mechanism}\label{sec_results_slsch}
Now, we evaluate the performance of a NR-V2X network with application of the proposed mechanism. We remind of two metrics for measurement of the performance, namely, BLER and normalized throughput. We also recall from Table \ref{table_parameters} that our focus is the SL-SCH for a groupcast or a unicast in NR-V2X mode 4.

Fig. \ref{fig_xi_mappedTBS} displays the four possible options for each of NPRB = \{6, 20\}. We selected from Table 7.1.7.2.1-1 of \cite{ts36213} \{152, 328, 712, 1032\} for NPRB of 6 and \{536, 1416, 2472, 3426\} for NPRB of 20 as an example. However, we reiterate that any other TBS value defined in the reference \cite{ts36213} could be eligible in the output space $\mathbf{y}$.

Figs. \ref{fig_xi_bler} and \ref{fig_xi_tput} show the resulting performance for each $x_{i}$ in terms of BLER and normalized throughput, respectively. The results commonly suggest that the proposed mechanism works as we intended in such a way that a higher crash-causing factor gets to yield a lower BLER. For instance, revisiting Table \ref{table_X}, a smaller index $i$ indicates a higher statistical gravity in causing a crash. Figs. \ref{fig_xi_bler} and \ref{fig_xi_tput} tell that our proposed mechanism leads a NR-V2X network to where $x$ with a smaller index $i$ achieves a lower BLER and a higher throughput. This way, a network can be managed in a way that \textit{a vehicle driven by a dangerously behaving driver can take the sidelink resource with a higher chance}, which will, in turn, elevates the chance of the air interface filled up with more urgent messages.

\section{Conclusions}\label{sec_conclusions}
\textit{Can we adapt multiple access for C-V2X according to the dynamically changing environment around a vehicle? For that, can a vehicle measure the crash risk around itself without support from infrastructure?} This paper laid out answers to these questions. Technically speaking, this paper presented a comprehensive algorithmic framework that features: (i) quantification of the driver's dangerous behaviors as the crash risk indicator of a vehicle; (ii) a contextual MAB algorithm for selection of an optimal TBS for SL-SCH in NR-V2X mode 4 adaptive to the driver's behavior; (iii) the algorithm's ability to operate at a vehicle autonomously without need for any support from a centralized entity. Indeed, our simulations found that the proposed mechanism was able to find an optimal TBS. This resulted in a more reliable performance (in terms of BLER and normalized throughput) for a more dangerously driven vehicle.

We identify as future work the ``relaxation'' of the regression of the driver's behavior and its crash causing statistics. While this paper characterized it as an order-12 polynomial kernel, it can be relaxed to a multi-kernel framework so it can accommodate a wider variety of driver behaviors in a more precise manner.


\end{document}